\documentclass{ws-procs975x65}
\usepackage{latexsym}

\begin{document}

\title{Nonlocality of Accelerated Systems}

\author{Bahram Mashhoon}

\address{Department of Physics and Astronomy, University of
Missouri-Columbia,\\Columbia, Missouri 65211, USA\\E-mail: 
mashhoonb@missouri.edu\\[\baselineskip] \textup{Dedicated to Prof. Remo Ruffini in honor of his 60th
birthday}}

\maketitle

\abstracts{The conceptual basis for the nonlocality of accelerated systems is
presented.  The nonlocal theory of accelerated observers and its consequences
are briefly described.   {\it Nonlocal field equations} are developed for the
case of the electrodynamics of linearly accelerated systems.}

\section{Introduction} Einstein's general relativity provides a satisfactory
description of gravitation at the classical level [1]; however, a theory of
gravitation valid at the quantum level has been elusive.  One way out of this
situation would be to return to the physical foundations of general relativity
and identify, from a physical standpoint, the main obstacle to quantization. 
The result of such an analysis [2-4] is that the {\it hypothesis of locality}
is the first major stumbling block.  This hypothesis provides the theoretical
connection between accelerated and inertial observers in the special theory of
relativity.

The fundamental quantum laws of physics have been formulated with respect to
ideal inertial observers, while actual observers are all noninertial.  It
follows from the success of the quantum theory that there must be a direct
connection between accelerated and inertial observers.  What is the nature of
this connection?  The standard theory of relativity postulates a {\it
pointwise} connection between actual and inertial observers.  It states that
an accelerated observer is equivalent, at each instant along its worldline, to
an otherwise identical momentarily comoving inertial observer.  This
hypothesis of locality amounts to replacing the worldline by its tangent
vector at each instant.  The idea originates from Newtonian mechanics of point
particles, where the accelerated particle and the momentarily comoving
inertial particle have the same {\it state} .  If all physical phenomena could
be described in terms of {\it pointlike coincidences}, then the hypothesis of
locality would be exactly valid.  

Lorentz invariance together with the hypothesis of locality constitutes the
physical basis of the special theory of relativity.  Moreover, the hypothesis
of locality provides the crucial link between the special and general theories
of relativity :  Einstein's heuristic principle of equivalence relates the
measurements of an observer in a gravitational field with those of an
accelerated observer in Minkowski spacetime; hence, this principle will lose
its operational significance if one does not know what accelerated observers
measure. 

This work is primarily concerned with the nonlocal measurements of noninertial
observers in Minkowski spacetime.  Therefore, imagine such an observer
following a worldline $x^{\mu}(\tau)$ in a global inertial frame with
coordinates $x^{\mu}=(ct, x, y, z)$.  Here $\tau$ is the proper time along the
worldline such that $-d\tau^2 = \eta_{\mu\nu}dx^{\mu}dx^{\nu}$, where
$\eta_{\mu\nu}$ is the Minkowski metric tensor with signature +2.  The
hypothesis of locality implies that the accelerated observer carries an
orthonormal tetrad frame $\lambda^{\mu}_{\:\:(\alpha)}$ such that
$\lambda^{\mu}_{\:\:(0)} = dx^{\mu}/d\tau$ is the temporal axis and
$\lambda^{\mu}_{\:\:(i)}, i = 1, 2, 3,$ are the spatial axes of the local frame
of the observer.  The {\it translational} acceleration is defined via
$A^{\mu} = d\lambda^{\mu}_{\:\:(0)}/d\tau$, while the acceleration of the
observer is given by the antisymmetric tensor $\Phi_{\alpha\beta}$,

\begin{equation}
\frac{d\lambda^{\mu}_{\:\:(\alpha)}}{d\tau} =
\Phi_{\alpha}^{\;\;\beta}\;\lambda^{\mu}_{\:\:(\beta)}\;\;.
\end{equation}

\noindent The ``electric'' part of the acceleration tensor $\Phi_{0i} = a_i$
characterizes the translational acceleration vector, i.e. $a_i =
A_{\mu}\lambda^{\mu}_{\:\:(i)}$, while the ``magnetic'' part of the
acceleration tensor $\Phi_{ij}=\epsilon_{ijk}\Omega^k$ characterizes the
rotational frequency $\mbox{\boldmath$\Omega$}$ of the local frame with
respect to a local nonrotating (i.e.\,Fermi-Walker transported) frame.  The
acceleration of the observer is thus completely characterized by the spacetime
scalars ${\bf a}(\tau)$ and $\mbox{\boldmath$\Omega$}(\tau)$, which determine
the local rate of change of the state of the observer.  To indicate the scale
of such variation, it is useful to define acceleration lengths ${\cal L} =
c^2/a$ and $c/\Omega$ as well as the corresponding acceleration times given by
${\cal L}/c$.  If $\lambda$ is the intrinsic length scale of the phenomenon
under observation, the deviation from the hypothesis of locality is expected
to be proportional to $\lambda/{\cal L}$.  If this deviation is so small as to
be below the detection capability of the observer, then the hypothesis of
locality is valid.  This is indeed the case for most Earth-bound experiments,
since
$c^2/g_{\oplus}\simeq 1$ lyr and $c/\Omega_{\oplus}\simeq 28$ AU.  As a matter
of principle, however, classical wave phenomena violate the hypothesis of
locality.  Consider, for instance, the reception of an electromagnetic wave by
the accelerated observer; to determine the properties of the incident wave,
the reception of a few oscillations are in any case necessary.  The hypothesis
of locality will hold if these oscillations have vanishing durations, i.e. in
the ray limit
$\lambda/{\cal L}\rightarrow 0$. 

There is an analog of the correspondence principle at work here:  The
hypothesis of locality is necessary for the establishment of the reference
frame of the observer as well as for the description of standard measuring
devices.  That is, inertial effects are always at work in any accelerated
measuring device.  If these inertial effects integrate to a measurable
influence over the duration of elementary local observations, then the device
may not be considered locally inertial.  To detect nonlocal effects, however,
one needs access to devices that are locally inertial.  

{\it Standard} measuring devices are those that are consistent with the
hypothesis of locality.  The basic spacetime measurements of inertial
observers involve ideal clocks and measuring rods; therefore, the
corresponding measurements of accelerated observers involve standard clocks
and standard measuring rods.  Thus standard clocks measure proper time and
standard infinitesimal measuring rods may be used to establish extended
coordinate systems in space.  The spatial determinations of ideal inertial
observers are in accordance with Euclidean geometry; therefore, in Minkowski
spacetime an accelerated observer at each instant along its worldline
experiences the Euclidean space in accordance with the hypothesis of
locality.  On this basis, a geodesic coordinate system may be established
along the worldline; however, this system naturally has spatial limitations
due to the requirement of the admissibility of the geodesic coordinates. 
Moreover, the conceptual consistency of spatial determinations by accelerated
observers has been critically examined by means of thought experiments with
the conclusion that finite standard measuring rods do not, in principle,
exist.  More precisely, it is not possible to put infinitesimal standard
measuring rods one next to another and so on in order to come up with a
definite length for the distance between two points in space.  In practice,
only distances that are negligibly small compared to the relevant acceleration
lengths may be unambiguously defined [2-5].  In view of these limitations,
accelerated reference frames are eschewed in the nonlocal theory of accelerated
observers.  Furthermore, severe restrictions must be placed on any spatial
dimension $D$ of a standard device; that is,
$D<<{\cal L}$, where ${\cal L}$ is the relevant acceleration scale.  These
considerations are consistent with the pointwise nature of the hypothesis of
locality.  

The nonlocal theory of accelerated observers is described in section 2. 
Nonlocal electrodynamics of linearly accelerated systems is considered in
section 3 and the nonlocal field equations for such observers are discussed in
section 4.  Section 5 contains a brief discussion.  For the sake of
simplicity, units are chosen such that $c=1$ in the rest of this paper.  

\section{Nonlocal theory of accelerated observers}

An accelerated observer along its worldline passes through a continuous
infinity of hypothetical momentarily comoving inertial observers.  The events
along the worldline can be uniquely identified by the proper time
$\tau$.  Let $\psi(x)$ be a radiation field in the background Minkowski
spacetime and $\hat\psi(\tau)$ be the field as measured by the hypothetical
inertial observer along the worldline at $\tau$.  At the same spacetime event,
${\hat\psi} = \Lambda\psi$, where $\Lambda$ belongs to a matrix representation
of the Lorentz group.  More precisely, suppose that
$x^{\prime} = Lx + s$ is a passive proper Poincar\'{e} transformation of the
inertial spacetime coordinates; then, $\psi^{\prime}(x^{\prime}) =
\Lambda(L)\psi(x)$, so that $\Lambda = 1$ for a scalar field.  Along the
worldline, events are characterized by the proper time $\tau$; hence,
${\hat\psi}(\tau) = \Lambda(\tau)\psi(\tau)$ is the measured field in
accordance with the hypothesis of locality.  

The most general linear relationship between $\Psi(\tau)$, which is the field
measured by the accelerated observer, and ${\hat\psi}(\tau)$ that is
consistent with the requirement of causality is 

\begin{equation}
\Psi(\tau) = {\hat\psi}(\tau) + \int^{\tau}_{\tau_0} K(\tau,
\tau^{\prime}){\hat\psi}(\tau^{\prime})d\tau^{\prime}\;\;,
\end{equation}

\noindent where $\tau_0$ is the instant at which the acceleration begins.  To
avoid unphysical situations, the accelerated motion is generally assumed to
have a finite duration.  Here $K(\tau,
\tau^{\prime})$ is a kernel that is directly related to the acceleration of
the observer, since in the absence of acceleration the kernel $K$ and hence
the nonlocal part of equation (2) must vanish.  Moreover, the nonlocal part is
expected to vanish for $\lambda/{\cal L}\rightarrow 0$ and equation (2) would
then reduce to the expression of the hypothesis of locality for field
measurement.  Thus equation (2) is consistent with the superposition principle
and involves a weighted average of the field over the past worldline of the
accelerated observer.  This  is in general agreement with the notion that
field measurement does not occur at a point but involves a certain spacetime
average [6, 7]. 

Equation (2) may be written as 

\begin{equation}
\Psi = \Lambda\psi+\int^{\tau}_{\tau_0} K(\tau,
\tau^{\prime})\Lambda(\tau^{\prime})\psi(\tau^{\prime})d\tau^{\prime}\;\;,
\end{equation} 

\noindent which is the general expression relating the field measured by the
static inertial observers $\psi$ to the field measured by accelerated
observers $\Psi$.  Equation (2) is a Volterra integral equation relating
$\hat\psi$ to $\Psi$.  According to Volterra's theorem, in the space of
continuous functions the relation between $\Psi$ and $\hat\psi$ (and hence
$\psi$) is unique [8].  This theorem has been extended to the Hilbert space of
square-integrable functions by Tricomi [9].  The Volterra-Tricomi uniqueness
result can be used in the determination of the kernel $K$. 

The kernel $K$ is determined on the basis of the postulate that a basic
radiation field can never stand completely still with respect to any observer
[10].  This is a generalization of the well-known fact that a basic radiation
field can never stand completely still with respect to an inertial observer. 
That is, in the Doppler formula $\omega^{\prime} =
\gamma\omega(1-{\bf v}\cdot{\bf\hat k})$, the only way that
$\omega^{\prime}$ can vanish is if $\omega$ vanishes.  This condition can be
implemented in the general case involving noninertial observers as well. 
Indeed, this requirement means that if $\Psi(\tau)$ should turn out to be a
constant, then $\psi(\tau)$ must have been a constant in the first place.  The
Volterra-Tricomi uniqueness theorem then implies that for any true radiation
field $\psi$, the measured field $\Psi$ will never be a constant.  Since
$\Psi(\tau_0) = {\hat\psi}(\tau_0) =
\Lambda(\tau_0)\psi(\tau_0)$, equation (3) satisfies the desired
condition---that if $\Psi$ is constant, then $\psi$ must be constant---once 

\begin{equation}
\Lambda(\tau_0) = \Lambda(\tau) + \int^{\tau}_{\tau_0}
K(\tau,\tau^{\prime})\Lambda(\tau^{\prime})d\tau^{\prime}
\end{equation}

\noindent holds.  A detailed examination of the solutions of equation (4) and
the comparison of the results with observational data on spin-rotation
coupling lead to the conclusion that the only acceptable kernel is of the form
[11]  

\begin{equation} K(\tau, \tau^{\prime}) = k(\tau^{\prime}) = -
\frac{d\Lambda(\tau^{\prime})}{d\tau^{\prime}}
\Lambda^{-1}(\tau^{\prime})\;\;.
\end{equation}

\noindent The kernel (5) is directly proportional to the acceleration of the
observer and is in fact a simple solution of equation (4) as can be verified
by direct substitution.  

The general nonlocal relation (3) with the kernel (5) is expected to hold in
the quantum domain as well, ensuring the independence of the number of the
quanta of the field from the motion of the observer. 

An immediate consequence of this nonlocal theory of accelerated systems may be
noted : the kernel vanishes for a constant $\Lambda$.  It follows from this
fact that the nonlocal theory forbids the existence of a pure scalar (or
pseudoscalar) radiation field.  Such a field must therefore be a composite. 
This circumstance is consistent with experimental results available at
present.  Further observational consequences of the nonlocal theory have been
discussed in [12].  

To illustrate the general theory described in this section, the rest of this
paper is devoted to nonlocal electrodynamics [13-15].  In particular, it is
the purpose of this paper to derive the corresponding nonlocal field equations
that would reduce to Maxwell's equations in the local limit. 

\section{Nonlocal electrodynamics}

Consider the measurement of the electromagnetic field by a linearly
accelerated observer.  That is, at $\tau = \tau_0$ the observer accelerates
from rest with acceleration $g(\tau)>0$ along the $z$-axis.  For
$\tau\geq\tau_0$, the orthonormal tetrad of the observer is given by 

\begin{eqnarray}
\lambda^{\mu}_{{\:\:}(0)} &=&(C\;,\; 0\;,\; 0\;,\; S)\;\;,\\
\lambda^{\mu}_{{\:\:}(1)} &=& (0\;,\; 1\;,\; 0\;,\; 0)\;\;,\\
\lambda^{\mu}_{{\:\:}(2)} &=& (0\;,\; 0\;,\; 1\;,\; 0)\;\;,\\
\lambda^{\mu}_{{\:\:}(3)} &=& (S\;,\; 0\;,\; 0\;,\; C)\;\;,
\end{eqnarray}

\noindent where $C = \cosh\theta, S = \sinh\theta$ and 

\begin{equation}
\theta = \int^{\tau}_{\tau_0} g(\tau^{\prime})d\tau^{\prime}\;\;.
\end{equation}

Let $F_{\mu\nu}$ be the Faraday tensor of an electromagnetic radiation field
in Minkowski spacetime.  At each instant of proper time $\tau$ along the
worldline, the momentarily comoving inertial observer measures

\begin{equation} {\hat F}_{\alpha\beta}(\tau) =
F_{\mu\nu}\lambda^{\mu}_{\:\:(\alpha)}\lambda^{\nu}_{\:\:(\beta)}\;\;.
\end{equation}

\noindent Using the natural decomposition $F_{\mu\nu}\rightarrow({\bf E}, {\bf
B})$, where $F_{0i} = -E_i$ and $F_{ij} = \epsilon_{ijk}B^k$, one can write
equation (11) as ${\hat F} = \Lambda F$, where $F$ is a column 6-vector with
${\bf E}$ and ${\bf B}$ as components and $\Lambda$ is a
$6\times 6$ matrix.  Imagine the measurement of electric and magnetic fields
via measuring devices carried by the accelerated observer.  If the
acceleration of the observer could be ignored over the length and time
scales of the measurement, then --- as shown by Bohr and Rosenfeld [6, 7]
for the case of ideal inertial observers --- charged particle dynamics in
accordance with the Lorentz force law would lead to the conclusion that one
would effectively measure the average field given in equation (11) with a
level of accuracy that would depend on the sensitivity of the measuring devices. 
On the other hand, the charged particles have been subject to the effects of
acceleration since $\tau_0$, as the observer and its comoving devices are
noninertial.  The nonlocality of field measurement therefore implies that equation (11)
should be considered a first approximation.  To separate the essential aspect
of this nonlocality from the limitations of measuring devices one needs the
most general nonlocal relationship between the actually measured field and the
field measured by the hypothetical momentarily comoving inertial observers. 
In this treatment, such a relationship is considered subject to the
superposition principle, causality and the postulate that a basic radiation
field would never stand completely still with respect to an observer.  Thus the
main equation of nonlocal electrodynamics is 

\begin{equation} {\hat{\cal F}}(\tau) = {\hat F}(\tau) +
\int^{\tau}_{\tau_0} k(\tau^{\prime}){\hat F}(\tau^{\prime})d\tau^{\prime}\;\;,
\end{equation}

\noindent where ${\hat{\cal F}}$ is the field measured by the accelerated
observer.  It is possible to write this equation in the form

\begin{equation} {\hat F}(\tau) = {\hat{\cal F}}(\tau) +
\int^{\tau}_{\tau_0}R(\tau,
\tau^{\prime}){\hat{\cal F}}(\tau^{\prime})d\tau^{\prime}\;\;,
\end{equation}

\noindent where $R(\tau, \tau^{\prime})$ is the resolvent kernel associated
with $k(\tau^{\prime})$. 

In the case under consideration, one can show that 

\begin{equation} 
\Lambda=\left[ \begin{array}{cc} U & V \\ -V & U
\end{array}\right]\;\;,\;\;U = \left[ \begin{array}{ccc} C & 0 & 0 \\ 0 & C &
0\\0 & 0 & 1\end{array}\right]\;\;,\;\; V = SI_3\;\;,
\end{equation}

\noindent where $I_i\;,\;(I_i)_{jk} = -\epsilon_{ijk}$, is a $3\times 3$ matrix
that is proportional to the operator of infinitesimal rotations about the
$x^i$-axis.  Here $UV=VU=CSI_3$ and $U^2 + V^2 = I$, where $I$ is the unit
matrix.  Using equation (5), the kernel $k$ is given by 

\begin{equation}  k=-g(\tau)\left[ \begin{array}{cc} 0 & I_3 \\ -I_3 & 0
\end{array}\right]\;\;.
\end{equation}

\noindent For the sake of simplicity, it will henceforth be assumed that
$g(\tau)$ is constant, $\tau_0=0$ and thus $\theta = g\tau$.  For a constant
kernel $k$, the resolvent kernel $R(\tau, \tau^{\prime})$ turns out to be of
the convolution (Faltung) type, $R(\tau, \tau^{\prime}) =
r(\tau-\tau^{\prime})$, where $r$ may be expressed as $r(u) = -k\;{\rm
exp}(-uk)$.  In general, $R(\tau, \tau^{\prime})$ could be very
complicated; indeed, the manageable expression for $r$ is due to the
assumption of uniform acceleration.  Therefore in the simple case under
consideration here the resolvent kernel turns out to be 

\begin{equation}  r(\tau)=g\left[ \begin{array}{cc} SJ_3 & CI_3 \\ -CI_3 & SJ_3
\end{array}\right]\;\;,
\end{equation}

\noindent where $(J_k)_{ij} = \delta_{ij} -
\delta_{ik}\delta_{jk}\;\;,\;\;J_3 = -I^{\;\;2}_3$ and $J_3I_3 = I_3J_3 =
I_3$.  It follows that equation (13) may be used to express the Faraday tensor
$F_{\mu\nu}$ in terms of the nonlocal electromagnetic field of a class of
accelerated observers.  Since $F_{\mu\nu}$ satisfies Maxwell's equations, one
can use equation (13) to derive the corresponding nonlocal field equations.

\section{Nonlocal Maxwell's equations}

Consider the class of fundamental inertial observers in the background global
system:  Each observer is at rest at an event $x^{\mu} = (t, {\bf x})$ and
carries an orthonormal frame whose axes coincide with those of the background
inertial system.  At $t=0$ the whole class is accelerated from rest along the
$z$-axis with uniform acceleration $g$.  Thus at any time $t>0$, an event $(t,
x, y, z)$ is occupied by an accelerated observer in hyperbolic motion that at
$t=0$ occupied $(0, x, y, z_0)$ such that 

\begin{equation} z_0 = z + \frac{1}{g} - \sqrt{t^2 + \frac{1}{g^2}}\;\;.
\end{equation}

\noindent At each such event $(t, x, y, z)$, one can define a field
${\cal F}(t, x, y, z)$ such that ${\hat{\cal F}}(\tau) = \Lambda{\cal F}$,
where $\tau$ is the proper time, $gt = \sinh g\tau$, along the hyperbolic
worldline.  This would correspond to defining a spacetime tensor ${\cal
F}_{\mu\nu}\rightarrow({\bf\mbox{\boldmath${\cal E}$}}, {\mbox{\boldmath${\cal
B}$}})$ such that 

\begin{equation} {\hat{\cal F}}_{\alpha\beta} = {\cal
F}_{\mu\nu}\lambda^{\mu}_{\:\:(\alpha)}\lambda^{\nu}_{\:\:(\beta)}\;\;.
\end{equation}

\noindent Equation (13) may now be written in the following general form 

\begin{equation} F(\tau) = {\cal F}(\tau) +
\int^{\tau}_{\tau_0} [\Lambda^{-1}(\tau)R(\tau,
\tau^{\prime})\Lambda(\tau^{\prime})] {\cal
F}(\tau^{\prime})d\tau^{\prime}\;\;.
\end{equation}

\noindent In the case under consideration one finds that 

\begin{equation}
\Lambda^{-1}(\tau)r(\tau-\tau^{\prime})\Lambda(\tau^{\prime}) = -k\;\;,
\end{equation}

\noindent where $\Lambda(u) = \exp(-uk)$, which follows from equation (5) and
the fact that $\Lambda(0)$ is the unit matrix.  Thus equation (19) takes the
form

\begin{equation} F(t, x, y, z) = {\cal F}(t, x, y, z) + g\,u_0(t)\left[
\begin{array}{cc} 0 & I_3 \\ -I_3 & 0\end{array}\right]\int^{\tau}_0{\cal
F}(t^{\prime}, x, y, z^{\prime})d\tau^{\prime}\;\;,
\end{equation}

\noindent where the unit step function $u_0(t), u_0 =1$ for $t>0$ and
$u_0=0$ otherwise, has been introduced to ensure nonlocality only for
$t>0$.  Using $\sinh g\tau = gt$, one can write $gd\tau = dt/\zeta(t)$, where 

\begin{equation}
\zeta(t) = \sqrt{t^2+\frac{1}{g^2}}\;\;.
\end{equation}

\noindent Moreover, in equation (21), $z^{\prime} = z_0 - g^{-1} +
\zeta(t^{\prime}),$ where $z_0 = z_0(t, z)$ is given by equation (17).  It
follows that 

\begin{equation} F(t, {\bf x}) = {\cal F}(t, {\bf x}) + u_0(t)\left[
\begin{array}{cc} 0 & I_3 \\ -I_3 & 0\end{array}\right]\int^t_0{\cal
F}(t^{\prime}, x, y, z -\zeta +
\zeta^{\prime})\frac{dt^{\prime}}{\zeta^{\prime}}\;\;,
\end{equation}

\noindent where $\zeta$ and $\zeta^{\prime}$ stand for $\zeta(t)$ and
$\zeta(t^{\prime})$, respectively.  This equation may be expressed as 

\begin{equation} {\bf E} = \mbox{\boldmath${\cal E}$} + u_0(t){\bf\hat
z}\;\times
\int^t_0\mbox{\boldmath${\cal B}$}(t^{\prime}, x, y,
z-\zeta+\zeta^{\prime})\frac{dt^{\prime}}{\zeta^{\prime}}\;\;,
\end{equation}

\begin{equation} {\bf B} = \mbox{\boldmath${\cal B}$} - u_0(t){\bf\hat
z}\;\times
\int^t_0\mbox{\boldmath${\cal E}$}(t^{\prime}, x, y,
z-\zeta+\zeta^{\prime})\frac{dt^{\prime}}{\zeta^{\prime}}\;\;,
\end{equation}

\noindent so that the fields are unchanged parallel to the direction of the
motion of the observer.

Introducing the Kramers vectors ${\bf W}^{\pm} = {\bf E}\pm i{\bf B}$, the
source-free Maxwell equations can be written in the form 

\begin{equation}
\mbox{\boldmath$\nabla$}\cdot{\bf W}^{\pm} = 0\;\;,
\end{equation}

\begin{equation}
\frac{1}{i}\mbox{\boldmath$\nabla$}\times{\bf W}^{\pm} =
\pm\frac{\partial}{\partial t}{\bf W}^{\pm}\;\;.
\end{equation}

\noindent When dealing with complex fields, ${\bf W}^+$ represents a
positive-helicity wave and ${\bf W}^-$ represents a negative-helicity wave. 
Equations (24) and (25) can be expressed as 

\begin{equation} {\bf W}^{\pm} = \mbox{\boldmath${\cal W}$}^{\pm}\mp
iu_0(t){\bf\hat z}\times
\int^t_0\mbox{\boldmath${\cal W}$}^{\pm}(t^{\prime}, x, y,
z-\zeta+\zeta^{\prime})\frac{dt^{\prime}}{\zeta^{\prime}}\;\;,
\end{equation}
where $\mbox{\boldmath$\mathcal{W}$}^{\pm} =
\mbox{\boldmath$\mathcal{E}$} \pm i \mbox{\boldmath$\mathcal{B}$}$.
Thus the nonlocal field equations are given by 

\begin{equation}
\mbox{\boldmath$\nabla$}\cdot\mbox{\boldmath${\cal W}$}^{\pm}=\mp 
iu_0(t){\bf\hat
z}\cdot\int^t_0\mbox{\boldmath$\nabla$}\times\mbox{\boldmath${\cal
W}$}^{\pm}(t^{\prime}, x, y, z-\zeta +
\zeta^{\prime})\frac{dt^{\prime}}{\zeta^{\prime}}\;\;,
\end{equation}

\begin{eqnarray}
\frac{1}{i}(\mbox{\boldmath$\nabla$} - \frac{u_0(t)}{\zeta}{\bf\hat z})
 \times 
\mbox{\boldmath${\cal W}$}^{\pm} = & \pm &\frac{\partial}{\partial
t}\mbox{\boldmath${\cal W}$}^{\pm}\nonumber \\
&\pm & u_0(t)\int^t_0\mbox{\boldmath$\nabla$}
\times ({\bf\hat z}  \times \mbox{\boldmath${\cal
W}$}^{\pm})(t^{\prime}, x, y,
z-\zeta+\zeta^{\prime})\frac{dt^{\prime}}{\zeta^{\prime}} \nonumber \\
& + & i\frac{tu_0(t)}{\zeta}{\bf\hat z}\times
\int^t_0\partial_z\mbox{\boldmath${\cal W}$}^{\pm}
(t^{\prime}, x, y, z-\zeta +
\zeta^{\prime})\frac{dt^{\prime}}{\zeta^{\prime}}\;\;.
\end{eqnarray}

\noindent Just as in equations (26) and (27), $\mbox{\boldmath${\cal
W}$}^+$ and  $\mbox{\boldmath${\cal W}$}^-$ satisfy separate equations, i.e.
nonlocality in general treats $\mbox{\boldmath${\cal W}$}^+$ and
$\mbox{\boldmath${\cal W}$}^-$ differently, but cannot convert one helicity
state into another.  Moreover, equations (26) and (27) can be combined in the
standard manner to yield the wave equations $\Box{\bf W}^\pm=0$.  A similar
procedure in the case of equations (29) and (30)---or, alternatively, applying
the d'Alembertian operator directly to equation (28)---results in rather
complicated and unwieldy wave equations for
$\mbox{\boldmath${\cal W}$}^\pm$.  

\section{Discussion}

Nonlocal field equations (29) and (30) have been derived in the background
global inertial frame.  These equations are Lorentz invariant, since they
originate from the manifestly Lorentz-invariant nonlocal ansatz (2). 
Moreover, it is possible to transform the nonlocal field equations to any
other coordinate system using the standard approach based on the invariance of
the 2-form ${\cal F}_{\mu\nu}\;dx^{\mu}\wedge dx^{\nu} = {\cal
F}^{\prime}_{\rho\sigma}\; dx^{\prime\rho}\wedge dx^{\prime\sigma}$.

The approach adopted in this paper for the development of nonlocal field
equations is quite general.  The resulting equations can be rather complicated,
however.  Thus attention has been confined to hyperbolic motion.  An infinite
amount of energy is required to sustain an observer in hyperbolic motion over
an infinite time interval; nevertheless, the case of uniform linear
acceleration has been treated explicitly in this work for the sake of
simplicity.

\end{document}